\documentclass[prl,twocolumn,superscriptaddress,showpacs,%
amsmath,amssymb,floatfix]{revtex4}
\usepackage{graphicx}

\newcommand{\Op}[1]{\boldsymbol{\mathsf{\hat{#1}}}}

\def\half{ \frac{1}{2}}
\def\openone{\leavevmode\hbox{\small1\kern-3.3pt\normalsize1}}

\begin{document}

\title{Creating Ground State Molecules with Optical Feshbach Resonances in Tight Traps}
\author{Christiane P. Koch}
\email{christiane.koch@lac.u-psud.fr}
\affiliation{Laboratoire Aim\'e Cotton, CNRS, B\^{a}t. 505, Campus d'Orsay,
91405 Orsay Cedex, France}
\affiliation{Department of Physical Chemistry and
The Fritz Haber Research Center, 
The Hebrew University, Jerusalem 91904, Israel}
\author{Fran\c{c}oise Masnou-Seeuws}
\affiliation{Laboratoire Aim\'e Cotton, CNRS, B\^{a}t. 505, Campus d'Orsay,
91405 Orsay Cedex, France}
\author{Ronnie Kosloff}
\affiliation{Department of Physical Chemistry and
The Fritz Haber Research Center, 
The Hebrew University, Jerusalem 91904, Israel}
\date{\today}

\begin{abstract}
We propose to create ultracold ground state molecules 
in an atomic Bose-Einstein condensate by adiabatic crossing
of an optical Feshbach resonance. We envision a scheme where the laser 
intensity and possibly also frequency are linearly ramped over the resonance.
Our calculations for $^{87}$Rb show that for sufficiently tight traps it is possible 
to avoid spontaneous emission while retaining adiabaticity, and conversion 
efficiencies of up to 50\% can be expected.
\end{abstract}
\pacs{32.80.Qk, 33.80.-b, 34.50.Rk}

\maketitle

The formation of ultracold molecules and the creation of molecular
Bose-Einstein condensates (BEC)~\cite{JochimSci03} open the way to  study 
new collective phenomena 
and a new, ultracold chemistry~\cite{RomPRL04,Chin04}. 
Since no direct cooling method for molecules  can reach the transition temperature for BEC, the formation
of molecules from ultracold atoms has been a focus of recent research.
Molecules are created by applying an external field,
either magnetic~\cite{DonleyNat02} or optical~\cite{McKenziePRL02},
to two colliding atoms.
This process is described in terms of a Feshbach
resonance (FR)~\cite{Feshbach} where the collision energy of the two atoms 
coincides with the energy of a bound molecular level.
Magnetic FR have been particularly successful in creating alkali dimer
molecules~\cite{DonleyNat02,StreckerPRL03},
even heteronuclear~\cite{StanPRL04}.
In contrast, optical FR involve electronically excited potentials, where
spontaneous emission may lead to loss of coherence~\cite{RomPRL04,McKenziePRL02}.
Apart from this obstacle, optical FR have the advantage that
optical transitions 
are almost always available, whereas magnetic FR require the presence of a
hyperfine manifold of the atom and 
may occur at magnetic field strengths which are difficult to obtain
in experiments. 
Furthermore,  optical FR offer more flexibility since two
parameters (laser intensity and frequency) instead of just one (magnetic
field strength) can be tuned. 
While optical FR have been employed to create molecules in cold gases
via photoassociation (PA)~\cite{FiorettiPRL98} and to tune the scattering 
length~\cite{TheisPRL04}, they have not yet been used to 
coherently create molecules except for the recent work of Ref.~\cite{RomPRL04}. 

\begin{figure}[b]
  \includegraphics[width=0.95\linewidth]{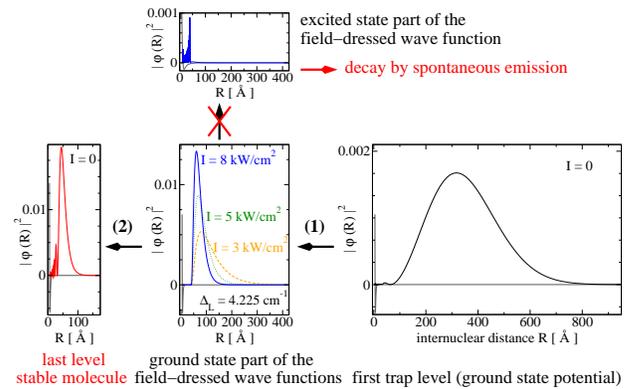}
   \caption{(Color online) Our proposed scheme for the creation of molecules: Step (1)
     is an adiabatic ramp of the laser intensity (and possibly frequency), 
     while step (2) is a sudden switch-off of the laser
     (wave functions shown for $\nu_\mathrm{tr}=50$~kHz).}
  \label{fig:pot}
\end{figure}

In this Letter, we propose to employ optical FR to 
create weakly bound ground state molecules (in singlet and
triplet ground state potentials both labelled 'ground state'
in the following). 
In analogy to magnetic FR, we envisage a scheme of adiabatically ramping 
over the resonance (cf. Fig.~\ref{fig:pot}). The resulting wave function
has components on both electronic ground and excited states with the latter
being subject to spontaneous emission losses. In a second
step, the laser field therefore needs to be switched off. This corresponds
to projecting the wave function onto the field-free eigenstates. 
The goal is to sweep
intensities and frequencies such that this projection is predominantly 
onto the last bound level of the ground state, i.e. onto stable molecules. 
Our scheme is different from one-color PA~\cite{WeinerRMP99}
which populates \textit{excited} state levels. It is rather similar in spirit 
to \textbf{Sti}mulated \textbf{R}aman \textbf{A}diabatic 
\textbf{P}assage~\cite{BergmannRMP98}
in that population of the excited state is minimized using a two-photon
transition. It differs from two-color PA~\cite{RomPRL04} since the
sudden switch-off breaks the symmetry of the coupling between the
bound molecular level and the trap (or continuum) state.
We show that for sufficiently tight traps, intensity and 
frequency of the field 
can be tuned such that spontaneous emission losses are avoided 
while adiabaticity is retained. Such ramps can be 
realized experimentally employing acousto-optical modulators or diode lasers.
Sufficiently tight confinement can be reached in  microscopic dipole 
traps~\cite{SchlosserNature01}
or deep optical lattices~\cite{GreinerNature02}. 

Our calculations are performed for $^{87}$Rb. 
The generality of the scheme is emphasized by employing both 
singlet and triplet ground state potentials.
We consider two $^{87}$Rb atoms which collide in an 
isotropic harmonic trap and interact with a continuous wave (CW) laser field. 
The center of mass motion is decoupled, and
the dynamics in the internuclear distance $R$ is governed by 
the Hamiltonian 
\begin{equation}
  \label{eq:H}
  \Op{H} = 
  \begin{pmatrix}
    \Op{H}_g && \hbar\Omega \\
     \hbar\Omega  && 
    \Op{H}_e - \hbar (\omega_0 -\Delta_L) - \frac{i\hbar}{2}\Gamma
  \end{pmatrix} \,.
\end{equation}
$\Op{H}_{g(e)} = \Op{T} + V_{g(e)}(\Op{R}) +(-)  V_\mathrm{tr}(\Op{R})$ 
is the single channel Hamiltonian with
$\Op{T}$ the kinetic energy operator and $V_{g(e)}(\Op{R})$
the ground (excited) state interaction potential. 
$V_\mathrm{tr}(\Op{R}) = \half m \omega_\mathrm{tr}^2 \Op{R}^2 $ is
the potential of the dipole trap, and $\Gamma$ the decay rate modelling spontaneous
emission. $m$ denotes the reduced mass and $\omega_\mathrm{tr}$ the
frequency of the trap ($\omega_\mathrm{tr}= 2\pi\times\nu_\mathrm{tr}$).
The frequency of the laser,
$\omega_L = \omega_0 -\Delta_L$ is  
red-detuned by $\Delta_L$ relative to the atomic resonance at $\omega_0$.
In Eq.~(\ref{eq:H}), we invoke the
dipole and rotating wave approximations (RWA). 
The Rabi frequency $\Omega$ is then given by 
$\Omega = E_0 \vec{D}(\Op{R})\cdot\vec{\epsilon} \approx E_0\vec{D}\cdot\vec{\epsilon}$,
where $E_0$ is the amplitude of the laser field, $\vec{D}(\Op{R})$ the dipole moment 
and $\vec{\epsilon}$ the polarization vector of the laser field.
$\vec{D}(\Op{R})\cdot\vec{\epsilon}$ is approximated
by its asymptotic value deduced from
standard long range calculations~\cite{DissMihaela}.
In Eq.~(\ref{eq:H}), we neglect the
hyperfine structure. This is justified for sufficiently detuning
the laser from the atomic resonances (about 4~cm$^{-1}$ or 120~GHz,
the largest energy difference between hyperfine levels is 7~GHz 
between $F=1$ and $F=2$ for 5$^2$S$_{1/2}$). 
The potentials $V_{g(e)}(\Op{R})$ have been obtained
by matching the results of \textit{ab initio} calculations~\cite{Park2001}
to the long-range dispersion potentials 
$V_\mathrm{asy}(\Op{R})= ( C_3/\Op{R}^3 +) C_6/\Op{R}^6+C_8/\Op{R}^8$.
The coefficients for the $5S+5S$ and $5S+5P$ asymptote
are respectively found in Ref.~\cite{MartePRL02} and Ref.~\cite{GuterresPRA02}.
The repulsive barrier of the 
ground state potentials has been adjusted to give a triplet (singlet) 
scattering length of 100~a$_0$ (90~a$_0$). 
The Hamiltonian, Eq.~(\ref{eq:H}), is represented on a grid, employing
a mapping procedure~\cite{SlavaJCP99} which reduces
 the number of required grid points by a factor of 5 to 30.

We proceed in two steps. First, we diagonalize the Hamiltonian, 
Eq.~(\ref{eq:H}), and obtain the field dressed eigenstates and eigenenergies 
as a function of laser intensity and frequency. The term $-i\hbar\Gamma/2$
causes the Hamiltonian to be non-Hermitian with complex
eigenvalues. $\Gamma$ is  assumed to be independent of $\Op{R}$
which is consistent with the approximation $\vec{D}(\Op{R})\approx \vec{D}$. 
Therefore the imaginary part of the eigenvalues becomes $\Gamma/2$ times
the projection of the eigenfunction onto the excited 
state~\footnote{%
Let's expand the eigenfunctions,
$|\varphi^\mathrm{dressed}_v\rangle$,
of the Hamiltonian, Eq.~(\ref{eq:H}), in the eigenbasis of the field-free 
Hamiltonian (with $\Gamma=0$), 
$|\varphi^\mathrm{g/e}_{v'}\rangle$.
The imaginary part of the eigenvalue is given by
$\langle \varphi^\mathrm{dressed}_v | 1/2 \Gamma \openone_e | 
\varphi^\mathrm{dressed}_v\rangle = 
1/2\Gamma \sum_{v'}|\langle\varphi^\mathrm{e}_{v'}|\varphi^\mathrm{dressed}_v\rangle|^2
=1/2\Gamma|\langle e|\varphi^\mathrm{dressed}_v\rangle|^2$.
We have verified this property numerically.
}.
In a second step,
we solve the time-dependent Schr\"odinger equation
 to illustrate the creation
of molecules. $\Gamma$ is then set equal to its asymptotic value, 
$\sqrt{2}\Gamma_\mathrm{at}$ with $\Gamma_\mathrm{at}=\hbar/\tau_\mathrm{at}$ and 
$\tau_\mathrm{at}(\mathrm{5S+5P}_{3/2})=26.24$~ns,
$\tau_\mathrm{at}(\mathrm{5S+5P}_{1/2})=27.70$~ns.

\begin{figure}[tbp]
  \includegraphics[width=0.9\linewidth]{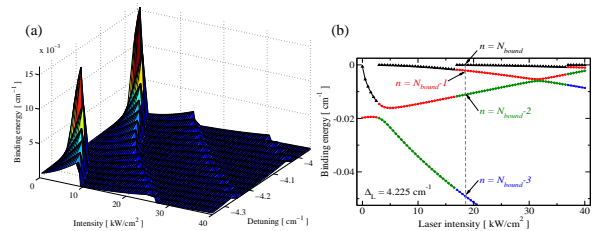}
  \caption{(Color online) (a) Binding energy of the last bound level below the
    (5S+5S) asymptote as a 
    function of laser intensity and detuning.
    (b) Binding energy of the last bound levels $n$ as a function of intensity. For $I $ =19~kW/cm$^2$ (vertical line) there are $N_{bound}$  levels below the (5S+5S) asymptote.    
  }
  \label{fig:E_3d}
\end{figure}

\begin{figure}[tbp]
   \includegraphics[width=0.9\linewidth]{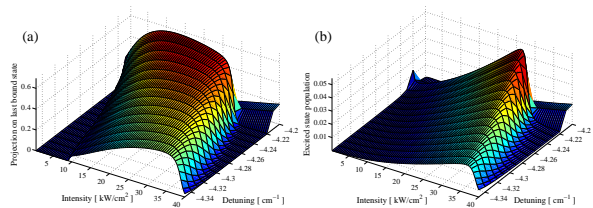}
  \caption{(Color online) Probability of molecule formation $P_\mathrm{mol}$ (a)
    and excited state component $p_\mathrm{exc}$ (b) as function
    of laser intensity and detuning
    ($\nu_\mathrm{tr}=250$~kHz).
  }
  \label{fig:proj}
\end{figure}

The following calculations are performed for transitions 
between the triplet ground state a$^3\Sigma_u^+$(5S+5S)
and the $0_g^-$(5S+5P$_{3/2}$) excited state. 
Fig.~\ref{fig:E_3d}a shows the binding energy of the last bound
level below the (5S+5S) asymptote
as a function of laser intensity and detuning. 
The range of detunings is chosen around 4~cm$^{-1}$, 
large enough to avoid hyperfine
coupling, and small enough such that the resonances
occur with excited state levels $v'$ which have  a good Franck-Condon overlap with
the last bound ground state level. 
Two resonances are found within this range 
($v'=40$ at 4.225~cm$^{-1}$ and $v'=41$ at 3.98~cm$^{-1}$). 
In Fig.~\ref{fig:E_3d}b,
the energies of the four last bound levels below the (5S+5S) asymptote
are plotted vs. laser intensity for a specific detuning.
Resonances at about 2.5~kW/cm$^2$, 16.5~kW/cm$^2$ and 36.5~kW/cm$^2$
are observed. At each resonance, the number of bound states is increased by
one. Usually, only the detuning is varied in optical FR. The increase
in the number of bound states can then be understood as 
follows: in the RWA ground and excited state potential cross and 
the excited state asymptote is at $\hbar\Delta_L$ above the ground
state dissociation limit. Decreasing the detuning therefore pushes
one more excited state level
below this dissociation limit.
The same happens as intensity is increased. 
It corresponds to the light shifts 
displacing the resonance positions with increasing intensity 
(cf. Fig.~\ref{fig:E_3d}a).
To further illustrate this "creation" of bound levels, 
Fig.~\ref{fig:pot} (middle) shows
the projection onto the ground state of one
field dressed wave function, 
$|\langle g |\varphi^\Omega_{n=81}\rangle|^2$ for different intensities, 
i.e. different $\Omega$ ($n$ counts all eigenstates).
At $I=0$, $|\varphi^{\Omega=0}_{n=81}\rangle$ coincides with the 
lowest trap state (Fig.~\ref{fig:pot}, right).
 As the intensity is increased, the wave function is deformed and pushed toward
shorter internuclear distances such that it eventually resembles the
wave function of the last bound level 
(Fig.~\ref{fig:pot}, left)~\footnote{Although the position $R_{max}$ of the outermost maximum
of the lowest trap state depends on the trap frequency, 
the intensities and
detunings at which the resonances occur are almost not affected by
the trap frequency. 
}.
The first step in our scheme  is therefore a slow ramp in intensity
(and possibly frequency) such that the wave function adiabatically
follows the field-dressed eigenfunctions, $|\varphi^\Omega_{n=81}\rangle$. 
In a second step, the field should  be suddenly switched off
projecting the field-dressed onto the field-free eigenfunctions. 
The probability to form a 
ground state molecule is then given by the 
projection of the field-dressed eigenfunction onto 
the last bound  ground state level, 
$P_\mathrm{mol}=|\langle\varphi^g_\mathrm{last} | \varphi^\Omega_{n=81}\rangle|^2$
(also lower bound levels can contribute to molecule
formation, but this is much less likely).
The lifetime of the  field-dressed eigenfunction,
 $\tau_v = \tau_\mathrm{at}/(\sqrt{2}p_\mathrm{exc})$, is determined 
by its excited state component, 
$p_\mathrm{exc}=|\langle e| \varphi^\Omega_{n=81}\rangle|^2$.
That is, $P_\mathrm{mol}$ corresponds to a gain while $p_\mathrm{exc}$ might
lead to a loss.
Both are shown in Fig.~\ref{fig:proj} 
as a function of laser intensity and detuning.
Close to resonance with an excited state level ($\Delta_L=-4.225$~cm$^{-1}$), 
at moderate intensities ($5$~kW/cm$^2 \le I\le 10$~kW/cm$^2$)
the projection onto the last bound level is 50\% and higher, 
while the excited state population does not exceed 0.01, i.e. the 
lifetime of the field-dressed eigenstate is  $\ge 2$~$\mu$s.
This lifetime defines an upper limit for the time window within which
the ramp across the resonance should be completed. The lower limit, $T_\mathrm{ad}$,
is due to the requirement of adiabaticity. It is determined by
the vibrational period of the lowest trap state, $T_\mathrm{vib}$,
which depends on the trap frequency and the interaction potential
(cf. Table~\ref{tab:Tvib}).
\begin{table}[btp]
  \begin{tabular}{|c|c|c|c|c|c|}
    \hline
    $\nu_\mathrm{tr}$ & 1~kHz & 50~kHz & 100~kHz & 250~kHz & 500~kHz \\ \hline
    $T_\mathrm{vib}$(a$^3\Sigma_u^+$) & 16.6~$\mu$s & 270~ns & 124~ns & 42.4~ns & 18.5~ns
    \\ \hline
    $T_\mathrm{vib}$($X^1\Sigma_g^+$) & 16.6~$\mu$s & 292~ns & 137~ns & 48.7~ns & 21.4~ns
    \\ \hline
  \end{tabular}
  \caption{Vibrational periods of the lowest trap state (calculated from its
  eigenenergy) for the triplet and singlet ground state potentials.}
  \label{tab:Tvib}
\end{table}
We can now estimate 
the timescales for our scheme:
Assuming the ramp should be performed in a time 
$T_\mathrm{ad} = 5\times T_\mathrm{vib}$ 
to be adiabatic,  spontaneous emission losses
should be minimal for a trap frequency
of $\nu_\mathrm{tr}\ge 250$~kHz ($T_\mathrm{ad}\approx 210$~ns). 
For $\nu_\mathrm{tr}\approx 50$~kHz, 
$T_\mathrm{ad}$ ($\approx 1.4$~$\mu$s) and $\tau_v$ are on the
same order of magnitude, and spontaneous emission losses will play a role.

\begin{figure}[tbp]
  \includegraphics[width=0.9\linewidth]{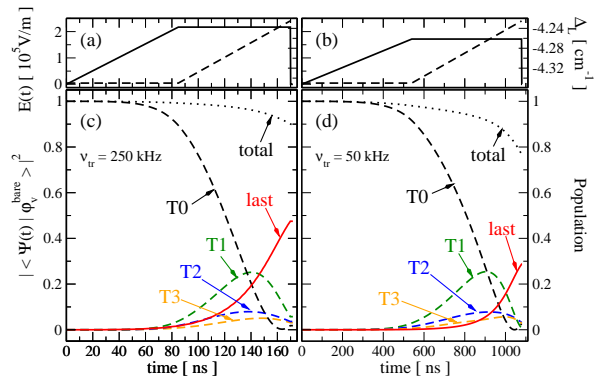}
  \caption{(Color online)
    (c)+(d): Projection of the time-dependent wave function
    onto the last bound ground state level (solid red line) and the first
    trap states $T$ (dashed lines) of the bare Hamiltonian. Also shown is the
    total population $|\langle\Psi(t)|\Psi(t)\rangle|^2$ (dotted line).
    (a)+(b): Variation of the laser field strength $E(t)$ (solid line) 
    and detuning $\Delta_L$ (dashed line). The maximum field intensity
    is $I=8$~kW/cm$^2$ for $\nu=250$~kHz and $I=5$~kW/cm$^2$ for $\nu=50$~kHz.
  }
  \label{fig:dyn}
\end{figure}

To verify our conclusions from the time-independent picture, 
we have explicitly studied the creation of molecules
solving
the time-dependent Schr\"odinger equation,
\begin{equation}
  \label{eq:TDSE}
  i\hbar \frac{\partial}{\partial t}|\Psi(t)\rangle = \Op{H}(t) |\Psi(t)\rangle \,,
\end{equation}
with a Chebychev propagator. The time-dependence 
in $\Op{H}(t)$ is due to the linear ramp in $\Omega$ (i.e. $E_0$ or $\sqrt{I}$) 
and $\omega_L$ (i.e. $\Delta_L$), respectively. Fig.~\ref{fig:dyn}
shows the projection of the wave function $\Psi(R;t)$ onto the last bound level
and onto the lowest trap levels of the 
field-free Hamiltonian vs. time.
Also plotted is the overall loss due to spontaneous emission ( 
we assume that any population undergoing spontaneous emission is lost 
from the coherence of the scheme, and possibly from the trap).
We combined a ramp in intensity 
with a ramp in frequency, both ramps are 
performed within 85~ns and 540~ns, respectively, 
i.e. $2\times T_\mathrm{vib}$. 
This turned out to be
sufficient to retain adiabaticity. While this result might be surprising 
at first glance, it reflects that the main source of nonadiabaticity is
Rabi cycling which in turn is suppressed by spontaneous emission. 
Simulations \textit{without} spontaneous emission showed that each ramp time had to be 
at least  $5\times T_\mathrm{vib}$ for the ramp to be adiabatic. 
Fig.~\ref{fig:dyn}c shows that for $\nu_\mathrm{tr}=250$~kHz,
almost 50\% of the population can be converted into ground state molecules, while
about 10\% are lost. The remaining population is distributed over the
lowest trap states. For  $\nu_\mathrm{tr}=50$~kHz
(Fig.~\ref{fig:dyn}d) the losses are somewhat higher at 24\%, 
but the conversion probability still reaches almost 30\%. 
Conversion probabilities higher than 50\% 
could be obtained for even tighter traps allowing for faster ramps.

We found that combining a ramp in intensity with a
ramp in frequency is the most efficient way to create molecules. The initial
ramp in intensity is performed with the laser frequency tuned in between
two resonances. Such a ramp deforms the wave function already considerably while
keeping the excited state population (cf. Fig.~\ref{fig:proj}b) and hence
the spontaneous emission loss  extremely small. 
In a second step the frequency is ramped toward the resonance. This ensures
a maximum overlap with the bound state wave function (cf. Fig.~\ref{fig:proj}a). 
Ramping up the intensity with the laser
tuned in between two resonances is not advantageous to create molecules. 
The overlap with the trap levels exceeds in this case by far the overlap 
with the bound state, i.e. one mainly excites higher trap states. 
If the intensity is ramped up with the laser tuned close to resonance,
due to the stronger coupling with the excited state, the ramp needs to be slower
to be adiabatic, and spontaneous emission losses are  larger
(cf. Fig.~\ref{fig:proj}).

Finally, since in the experiment the laser  cannot be switched off 
instantly, we address the question of a finite switch-off time after the ramp.
(cf. Table~\ref{tab:switch}).
\begin{table}[btp]
  \begin{tabular}{|l|c|c|c|c|}
    \hline
    $T_\mathrm{switch}$     & 0       & 10 ns   & 5 ns  & 1 ns \\ \hline
    $P$ with $\nu=250$~kHz  & 0.475   & 0.269   & 0.389 & 0.466  \\
    $P$ with $\nu=50$~kHz   & 0.288   & 0.203   & 0.246 & 0.284  \\
    \hline
  \end{tabular}
  \caption{Final probability of ground state molecule formation, 
    $P=|\langle \Psi(t_\mathrm{final}) | \varphi_g^\mathrm{bare}\rangle|^2$,
  for different switch-off times of the laser}
  \label{tab:switch}
\end{table}
As can be seen in Table~\ref{tab:switch},
 the field should ideally be switched off
within 1~ns. The probability of molecule formation is then
reduced by less than 1\%.
Switching off the field within 10~ns causes about one third 
of the created molecules to be lost, i.e. to be transferred
back into a pair of  trapped atoms. Due to the shorter timescales,
this problem is more severe in tighter traps.

We have performed the same set of calculations for a transition between the
X$^1\Sigma_g^+$(5S+5S) and $0_u^+$(5S+5P$_{1/2}$) potentials, which differ markedly from the previous ones. The common 
point is the  asymptotic $1/R^3$ behaviour in the excited  $0_u^+$(5S+5P$_{1/2}$) potential, also providing extremely long-range levels with a large Franck-Condon
overlap with the lowest trap state(s).
We therefore find exactly the same pattern of binding energies, 
probability of molecule formation and excited state population as 
shown in Figs.~\ref{fig:E_3d}-\ref{fig:dyn}~\footnote{%
Due to larger Franck-Condon factors, the resonances occur at 
 lower intensities for the same detuning of  4~cm$^{-1}$. 
}. 

To summarize we have shown that in tight traps,
loosely bound ground state molecules can be created 
efficiently and  without loss of coherence
by adiabatic ramping over an optical FR.
Both  detuning and intensity are varied within 
an asymmetric scheme which involves first adiabatic following and then
a fast switch-off. Three-body effects are neglected in our model: 
it is therefore applicable to low densities or ideally to a
Mott-Insulator state~\cite{GreinerNature02}.
We assumed the trap to be isotropic. However we expect our scheme
to work also for anisotropic traps as long as no new timescale is introduced, 
i.e. as long as the lowest trap frequency is $\gtrsim$ 50~kHz.
We point out that in shallow traps the  creation of molecules 
by optical FR does not seem to be feasible. In that case
 non-adiabatic processes or adiabatic schemes 
employing short laser pulses~\cite{ElianePRA04} should be considered.
For tight traps by contrast, calculations performed for rubidium atoms demonstrated
conversion probabilites up to 50\%. The presented scheme  should work for any optical 
transition which is also efficient in photoassociation, involving long-range
wells or  resonant
coupling in the excited state~\cite{ClaudePRL01}. The range of possible
applications therefore extends well beyond
homonuclear systems and alkali atoms. 

\begin{acknowledgments}
  We would like to thank P. Naidon, P. Grangier, J. Hecker Denschlag and R. Grimm
  for very fruitful discussions.
  This work has been supported by the EC in the frame of the 
  Cold Molecule network (contract HPRN-CT-2002-00290).
  C.P.K. acknowledges financial support from the Deutsche Forschungsgemeinschaft.
  The Fritz Haber Center is supported
  by the Minerva Gesellschaft f\"{u}r die Forschung GmbH M\"{u}nchen, Germany.
\end{acknowledgments}


\begin{thebibliography}{33}
\expandafter\ifx\csname natexlab\endcsname\relax\def\natexlab#1{#1}\fi
\expandafter\ifx\csname bibnamefont\endcsname\relax
  \def\bibnamefont#1{#1}\fi
\expandafter\ifx\csname bibfnamefont\endcsname\relax
  \def\bibfnamefont#1{#1}\fi
\expandafter\ifx\csname citenamefont\endcsname\relax
  \def\citenamefont#1{#1}\fi
\expandafter\ifx\csname url\endcsname\relax
  \def\url#1{\texttt{#1}}\fi
\expandafter\ifx\csname urlprefix\endcsname\relax\def\urlprefix{URL }\fi
\providecommand{\bibinfo}[2]{#2}
\providecommand{\eprint}[2][]{\url{#2}}

\bibitem[{\citenamefont{Jochim et~al.}(2003)\citenamefont{Jochim, Bartenstein,
  Altmeyer, Hendl, Riedl, Chin, Hecker~Denschlag, and Grimm}}]{JochimSci03}
\bibinfo{author}{\bibfnamefont{S.}~\bibnamefont{Jochim et~al.}},
  \bibinfo{journal}{Science} \textbf{\bibinfo{volume}{302}},
  \bibinfo{pages}{2101} (\bibinfo{year}{2003}).
%
  \bibinfo{author}{\bibfnamefont{M.}~\bibnamefont{Greiner}},
  \bibinfo{author}{\bibfnamefont{C.~A.} \bibnamefont{Regal}}, \bibnamefont{and}
  \bibinfo{author}{\bibfnamefont{D.~S.} \bibnamefont{Jin}},
  \bibinfo{journal}{Nature} \textbf{\bibinfo{volume}{426}},
  \bibinfo{pages}{537} (\bibinfo{year}{2003}).
%
  \bibinfo{author}{\bibfnamefont{M.~W.} \bibnamefont{Zwierlein et~al.}},
  \bibinfo{journal}{Phys. Rev. Lett.} \textbf{\bibinfo{volume}{91}},
  \bibinfo{pages}{250401} (\bibinfo{year}{2003}).

\bibitem[{\citenamefont{Rom et~al.}(2003)\citenamefont{Rom, Best, Mandel,
  Widera, Greiner, H\"ansch and Bloch}}]{RomPRL04}
  \bibinfo{author}{\bibfnamefont{T.} \bibnamefont{Rom et~al.}},
  \bibinfo{journal}{Phys. Rev. Lett.} \textbf{\bibinfo{volume}{93}},
  \bibinfo{pages}{073002} (\bibinfo{year}{2004}).

\bibitem[{\citenamefont{Chin et~al.}(2004)\citenamefont{Chin, Kraemer, Mark,
  Herbig, Waldburger, N\"agerl, and Grimm}}]{Chin04}
\bibinfo{author}{\bibfnamefont{C.}~\bibnamefont{Chin et~al.}},
  \bibinfo{journal}{cond-mat/0411258}  (\bibinfo{year}{2004}).

\bibitem[{\citenamefont{Donley et~al.}(2002)\citenamefont{Donley, Claussen,
  Thompson, and Wieman}}]{DonleyNat02}
\bibinfo{author}{\bibfnamefont{E.~A.} \bibnamefont{Donley et~al.}},
\bibinfo{journal}{Nature}
  \textbf{\bibinfo{volume}{417}}, \bibinfo{pages}{529} (\bibinfo{year}{2002}).

\bibitem[{\citenamefont{McKenzie et~al.}(2002)\citenamefont{McKenzie,
  Hecker~Denschlag, H\"affner, Browaeys, de~Araujo, Fatemi, Jones, Simsarian,
  Cho, Simoni et~al.}}]{McKenziePRL02}
\bibinfo{author}{\bibfnamefont{C.}~\bibnamefont{McKenzie et~al.}},
  \bibnamefont{et~al.}, \bibinfo{journal}{Phys. Rev. Lett.}
  \textbf{\bibinfo{volume}{88}}, \bibinfo{pages}{120403}
  (\bibinfo{year}{2002}).

\bibitem[{\citenamefont{Feshbach}(1958)}]{Feshbach}
\bibinfo{author}{\bibfnamefont{H.}~\bibnamefont{Feshbach}},
  \bibinfo{journal}{Ann. Phys.} \textbf{\bibinfo{volume}{5}},
  \bibinfo{pages}{357} (\bibinfo{year}{1958}).

\bibitem[{\citenamefont{Strecker et~al.}(2003)\citenamefont{Strecker,
  Partridge, and Hulet}}]{StreckerPRL03}
\bibinfo{author}{\bibfnamefont{K.~E.} \bibnamefont{Strecker}},
  \bibinfo{author}{\bibfnamefont{G.~B.} \bibnamefont{Partridge}},
  \bibnamefont{and} \bibinfo{author}{\bibfnamefont{R.~G.} \bibnamefont{Hulet}},
  \bibinfo{journal}{Phys. Rev. Lett.} \textbf{\bibinfo{volume}{91}},
  \bibinfo{pages}{080406} (\bibinfo{year}{2003}).
%
  \bibinfo{author}{\bibfnamefont{K.}~\bibnamefont{Xu et~al.}},
  \bibinfo{journal}{Phys. Rev. Lett.} \textbf{\bibinfo{volume}{91}},
  \bibinfo{pages}{210402} (\bibinfo{year}{2003}).
%
  \bibinfo{author}{\bibfnamefont{C.~A.} \bibnamefont{Regal et~al.}},
  \bibinfo{journal}{Nature} \textbf{\bibinfo{volume}{424}}, \bibinfo{pages}{47}
  (\bibinfo{year}{2003}).
%
\bibinfo{author}{\bibfnamefont{S.}~\bibnamefont{D\"urr et~al}},
  \bibinfo{journal}{Phys. Rev. Lett.} \textbf{\bibinfo{volume}{92}},
  \bibinfo{pages}{020406} (\bibinfo{year}{2004}).
%
  \bibinfo{author}{\bibfnamefont{J.}~\bibnamefont{Herbig et~al.}},
  \bibinfo{journal}{Science} \textbf{\bibinfo{volume}{301}},
  \bibinfo{pages}{1510} (\bibinfo{year}{2003}).


\bibitem[{\citenamefont{Stan et~al.}(2004)\citenamefont{Stan, Zwierlein,
  Schunck, Raupach, and Ketterle}}]{StanPRL04}
\bibinfo{author}{\bibfnamefont{C.~A.} \bibnamefont{Stan et~al.}},
  \bibinfo{journal}{Phys. Rev. Lett.} \textbf{\bibinfo{volume}{93}},
  \bibinfo{pages}{143001} (\bibinfo{year}{2004}).
%
\bibinfo{author}{\bibfnamefont{S.}~\bibnamefont{Inouye et~al.}},
  \bibinfo{journal}{Phys. Rev. Lett.} \textbf{\bibinfo{volume}{93}},
  \bibinfo{pages}{183201} (\bibinfo{year}{2004}).

\bibitem[{\citenamefont{Fedichev et~al.}(1996)\citenamefont{Fedichev, Kagan,
  Shlyapnikov, and Walraven}}]{FedichevPRL96}
\bibinfo{author}{\bibfnamefont{P.~O.} \bibnamefont{Fedichev et~al.}},
\bibinfo{journal}{Phys. Rev. Lett.}
  \textbf{\bibinfo{volume}{77}}, \bibinfo{pages}{2913} (\bibinfo{year}{1996}).
%
%

\bibitem[{\citenamefont{Fioretti et~al.}(1998)\citenamefont{Fioretti, Comparat
 Crubellier, Dulieu, Masnou-Seeuws and Pillet}}]{FiorettiPRL98}
\bibinfo{author}{\bibfnamefont{A.}~\bibnamefont{Fioretti et~al.}},
  \bibinfo{journal}{Phys. Rev. Lett.} \textbf{\bibinfo{volume}{80}},
  \bibinfo{pages}{4402} (\bibinfo{year}{1998}).


\bibitem[{\citenamefont{Theis et~al.}(2004)\citenamefont{Theis, Thalhammer,
  Winkler, Hellwig, Ruff, Grimm, and Hecker~Denschlag}}]{TheisPRL04}
\bibinfo{author}{\bibfnamefont{M.}~\bibnamefont{Theis et~al.}},
  \bibinfo{journal}{Phys. Rev. Lett.} \textbf{\bibinfo{volume}{93}},
  \bibinfo{pages}{123001} (\bibinfo{year}{2004}).

\bibitem[{\citenamefont{Weiner et~al.}(1999)\citenamefont{Weiner, Bagnato,
  Zilio, and Julienne}}]{WeinerRMP99}
\bibinfo{author}{\bibfnamefont{J.}~\bibnamefont{Weiner et~al.}},
  \bibinfo{journal}{Rev. Mod. Phys.} \textbf{\bibinfo{volume}{71}},
  \bibinfo{pages}{1} (\bibinfo{year}{1999}).

\bibitem[{\citenamefont{Bergmann et~al.}(1998)\citenamefont{Bergmann, Theuer,
  and Shore}}]{BergmannRMP98}
\bibinfo{author}{\bibfnamefont{K.}~\bibnamefont{Bergmann}},
  \bibinfo{author}{\bibfnamefont{H.}~\bibnamefont{Theuer}}, \bibnamefont{and}
  \bibinfo{author}{\bibfnamefont{B.~W.} \bibnamefont{Shore}},
  \bibinfo{journal}{Rev. Mod. Phys.} \textbf{\bibinfo{volume}{70}},
  \bibinfo{pages}{1003} (\bibinfo{year}{1998}).

\bibitem[{\citenamefont{Schlosser et~al.}(2001)\citenamefont{Schlosser,
  Reymond, Protsenko, and Grangier}}]{SchlosserNature01}
\bibinfo{author}{\bibfnamefont{N.}~\bibnamefont{Schlosser et~al.}},
  \bibinfo{journal}{Nature} \textbf{\bibinfo{volume}{411}},
  \bibinfo{pages}{1024} (\bibinfo{year}{2001}).
%
\bibinfo{author}{\bibfnamefont{S.}~\bibnamefont{Kuhr et~al.}},
  \bibinfo{journal}{Science} \textbf{\bibinfo{volume}{293}},
  \bibinfo{pages}{278} (\bibinfo{year}{2001}).

\bibitem[{\citenamefont{Greiner et~al.}(2002)\citenamefont{Greiner, Mandel,
  Esslinger, H\"ansch, and Bloch}}]{GreinerNature02}
\bibinfo{author}{\bibfnamefont{M.}~\bibnamefont{Greiner et~al.}},
  \bibinfo{journal}{Nature} \textbf{\bibinfo{volume}{415}}, \bibinfo{pages}{39}
  (\bibinfo{year}{2002}).

\bibitem[{\citenamefont{Vatasescu}(1999)}]{DissMihaela}
\bibinfo{author}{\bibfnamefont{M.}~\bibnamefont{Vatasescu}}, Ph.D. thesis,
  \bibinfo{school}{Universit\'{e} Paris XI}
  (\bibinfo{year}{1999}).

\bibitem[{\citenamefont{Park et~al.}(2001)\citenamefont{Park, Suh, Lee, and
  Jeung}}]{Park2001}
\bibinfo{author}{\bibfnamefont{S.~J.} \bibnamefont{Park et~al.}},
  \bibinfo{journal}{J. Molec. Spec.} \textbf{\bibinfo{volume}{207}},
  \bibinfo{pages}{129} (\bibinfo{year}{2001}).

\bibitem[{\citenamefont{Marte et~al.}(2002)\citenamefont{Marte, Volz, Schuster,
  D\"urr, Rempe, van Kempen, and Verhaar}}]{MartePRL02}
\bibinfo{author}{\bibfnamefont{A.}~\bibnamefont{Marte et~al.}},
 \bibinfo{journal}{Phys. Rev. Lett.}
  \textbf{\bibinfo{volume}{89}}, \bibinfo{pages}{283202}
  (\bibinfo{year}{2002}).

\bibitem[{\citenamefont{Gutteres et~al.}(2002)\citenamefont{Gutteres, Amiot,
  Fioretti, Gabbanini, Mazzoni, and Dulieu}}]{GuterresPRA02}
\bibinfo{author}{\bibfnamefont{R.~F.}~\bibnamefont{Gutterres et~al.}},
  \bibinfo{journal}{Phys. Rev. A} \textbf{\bibinfo{volume}{66}},
  \bibinfo{pages}{024502} (\bibinfo{year}{2002}).

\bibitem[{\citenamefont{Kokoouline et~al.}(1999)\citenamefont{Kokoouline,
  Dulieu, Kosloff, and Masnou-Seeuws}}]{SlavaJCP99}
\bibinfo{author}{\bibfnamefont{V.}~\bibnamefont{Kokoouline et~al.}},
  \bibinfo{journal}{J. Chem. Phys.} \textbf{\bibinfo{volume}{110}},
  \bibinfo{pages}{9865} (\bibinfo{year}{1999}).
%
\bibinfo{author}{\bibfnamefont{K.}~\bibnamefont{Willner}},
  \bibinfo{author}{\bibfnamefont{O.}~\bibnamefont{Dulieu}}, \bibnamefont{and}
  \bibinfo{author}{\bibfnamefont{F.}~\bibnamefont{Masnou-Seeuws}},
  \bibinfo{journal}{J. Chem. Phys.} \textbf{\bibinfo{volume}{120}},
  \bibinfo{pages}{548} (\bibinfo{year}{2004}).

\bibitem[{\citenamefont{Dion et~al.}(2001)\citenamefont{Dion, Drag, Dulieu,
  Laburthe~Tolra, Masnou-Seeuws, and Pillet}}]{ClaudePRL01}
\bibinfo{author}{\bibfnamefont{C.~M.} \bibnamefont{Dion et~al.}},
  \bibinfo{journal}{Phys. Rev. Lett.} \textbf{\bibinfo{volume}{86}},
  \bibinfo{pages}{2253} (\bibinfo{year}{2001}).
%
\bibinfo{author}{\bibfnamefont{O.}~\bibnamefont{Dulieu}} \bibnamefont{and}
  \bibinfo{author}{\bibfnamefont{F.}~\bibnamefont{Masnou-Seeuws}},
  \bibinfo{journal}{J. Opt. Soc. Am. B} \textbf{\bibinfo{volume}{20}},
  \bibinfo{pages}{1083} (\bibinfo{year}{2003}).

\bibitem[{\citenamefont{Luc-Koenig
  et~al.}(2004{\natexlab{a}})\citenamefont{Luc-Koenig, Kosloff, Masnou-Seeuws,
  and Vatasescu}}]{ElianePRA04}
\bibinfo{author}{\bibfnamefont{E.}~\bibnamefont{Luc-Koenig et~al.}},
  \bibinfo{journal}{Phys. Rev. A} \textbf{\bibinfo{volume}{70}},
  \bibinfo{pages}{033414} (\bibinfo{year}{2004}{\natexlab{a}}).
%
\bibinfo{author}{\bibfnamefont{E.}~\bibnamefont{Luc-Koenig et~al }},
  \bibinfo{journal}{Eur. Phys. J. D} \textbf{\bibinfo{volume}{31}},
  \bibinfo{pages}{239} (\bibinfo{year}{2004}{\natexlab{b}}).

\end{thebibliography}

\end{document}